\documentclass[prb,aps,twocolumn,showpacs,preprintnumbers,amsmath,amssymb]{revtex4}

\usepackage{graphicx}
\usepackage{dcolumn}
\usepackage{bm}

\begin{document}


\title{Spreading of Block Copolymer Films and Domain Alignment at Moving Terrace Steps\footnote{The following article has been submitted to the Journal of Chemical Physics. After it is published, it will be found at http://ojps.aip.org/jcpo}}

\author{Vladimir A. Belyi}
 \email{v-belyi@uchicago.edu}
\author{Thomas A. Witten}%
\affiliation{%
James Franck Institute and the Department of Physics, University of Chicago \\
5640 S. Ellis Avenue, Chicago, Illinois 60637
}

\date{\today}

\newcommand{\ie}{\textit{i.e. }}
\newcommand{\erf}{{\rm erf }}
\newcommand{\erfc}{{\rm erfc }}
\newcommand{\erfi}{{\rm erfi }}

\begin{abstract}
We investigate spreading of phase separated copolymer films, where domain walls and thickness steps influence polymer flow. We show that at early stages of spreading its rate is determined by slow activated flow at terrace steps (\ie thickness steps). At late stages of spreading, on the other hand, the rate is determined by the flow along terraces, with diffusion-like time dependence $t^{-1/2}$. This dependence is similar to de Gennes and Cazabat's prediction for generic layered liquids \cite{Cazabat}, as opposed to the classical Tanner's law of drop spreading. We also argue that chain hopping at the spreading terrace steps should lead to the formation of aligned, defect-free domain patterns on the growing terraces.
\end{abstract}


\maketitle

\section{\label{secIntroduction}Introduction}

Equilibrium structure of block copolymer systems was in the center of theoretical and experimental research for a number of years now, and is mostly understood \cite{BatesFredrickson1996,BatesFredrickson1990,Krausch,Fasolka,HarrisonInScience,HahmDefects,Morkved,Hahm}. It is known that in the bulk, at temperatures below order-disorder transition (ODT), molten diblock copolymers phase-separate into periodic structures with periodicity dependent on polymer size. The actual structures formed depend upon the proportion of different blocks inside copolymer, and include lamellar, cylindrical, spherical and various bicontinuous morphologies. 

The spreading dynamics of copolymer films, on the other hand, is much less investigated \cite{Joanny1995}. While little difference besides significantly higher viscosity is expected in the spreading of macroscopic copolymer drop, large deviations may arise on microscopic length scales, when the film thickness approaches domain period. In particular, thin films inherit domain structure of the bulk samples, but their domains are typically aligned parallel or perpendicular to the substrate. Combined with extreme anisotropy of diffusion relative to the domain orientation, spreading dynamics of thin films is expected to be very different from that of simple fluids.

\begin{figure}[tb]
\centerline{\includegraphics[width=2.4in,height=1.5in] {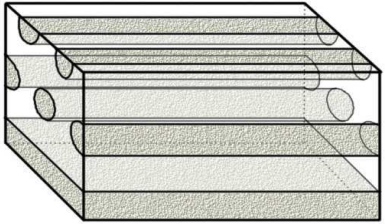}}
\caption{Sketch of the layered structure of the cylinder forming PS-PMMA copolymer discussed in the text. Shaded regions correspond to PMMA domains.}
\label{figLayers}
\end{figure}

In the present paper we look at the spreading of phase-separated copolymer films on wetting substrates in the regime of complete wetting. It is known that copolymer films at small thicknesses adopt a terraced topology, with thickness of each terrace defined by the underlying domain structure \cite{Coulon}. The copolymer domains are then localized inside these terraces (Fig. \ref{figLayers}), so that terraces can slide on top of each other. 

General spreading of layered drops was first investigated by de Gennes and Cazabat \cite{Cazabat,Voue2001}, and consists of viscous flow along each terrace, combined with disjoining pressure driven flow at terrace steps. Here we investigate spreading dynamics of the layered copolymer films. On one hand, the thickness of copolymer terraces (layers) significantly exceeds atomic sizes, so that effect of disjoining pressure is negligible at all but the bottommost terrace. Hence, the flow between terraces is driven by other factors, presumably by jump in the internal stress. Additionally, the thickness of each copolymer terrace needs not be uniform. This results in the forced diffusive flow of material along the terrace with diffusion constant determined by the underlying domain topology and strength of copolymer block interaction. We find that at large times the spreading rate $v$ for the copolymer film takes the diffusion like time dependence $v \propto t^{-1/2}$, as in general layered drops of Refs. \onlinecite{Cazabat} and \onlinecite{Voue2001}, and contrary to the Tanner's law for simple liquids, which predicts $v \propto t^{-0.9}$.

The activated flow of chains at terrace steps, with chains hopping between copolymer domains, can also account for the growth of aligned domain patterns in spreading films. Specifically, asymmetric copolymers that phase separate into cylindrical domains in bulk, often form striped pattern in the thin films, similar to those shown on Figure \ref{figLayers} \cite{Jaeger,HarrisonInScience,Morkved}. Easy control over periodicity of these patterns has driven large interest in their formation, but lack of ordering and generally large number of defects limited their applicability \cite{HarrisonInScience,HahmDefects,Morkved,Hahm}. 

\begin{figure}[tb]
\centerline{\includegraphics[width=2.4in,height=1.5in]{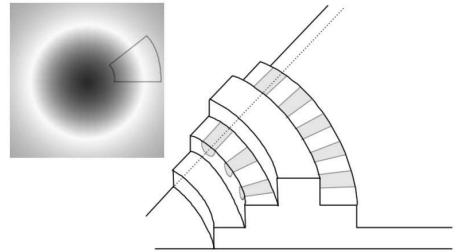}}
\caption{Sketch of annular structures observed by Hahm and Sibener, based on Figs. 2 and 3 of Ref. \onlinecite{Hahm}. Nearly perfect alignment of cylindrical domains in the radial direction was observed on the top surfaces of the terraces. In the actual experiment terraces extension in radial direction is 50 times larger than their thickness.}
\label{figHahm}
\end{figure}

However, Hahm and Sibener \cite{Hahm} recently reported that nearly perfect alignment is observed on annuli structures of polystyrene-block-polymethylmethacrylate (PS-PMMA). In that experiment, a thin film of PS-PMMA was spin cast onto a SiO$_2$ substrate that was initially covered with a minor polar solvent. Dewetting properties of the solvent resulted in the formation of bare circular regions with thick annular rims (Fig. \ref{figHahm}). On subsequent annealing the thick rims spread out and flatten. The expected striped domains showed unexpected radial alignment. This strong alignment along the spreading direction was confirmed in later experiments with a different geometry \cite{Deepak}. Below we will argue that hopping of copolymer chains between domains at terrace steps, necessary for the spreading, may be responsible for this alignment. In particular, chain hopping should lead to the continuous growth of cylindrical domains, while thickness gradient near terrace steps should generate alignment of the growing domains. As a result, one may expect aligned, defect-free domain patterns along the whole area of spreading, similar to the observations by Hahm and Sibener \cite{Hahm}.

Improving the ordering of copolymer film domains is regarded as a key to many potential applications \cite{Zehner,Mansky,Harrison1998b}. Accordingly, much attention has been given to potential alignment mechanisms: local alignment near a terrace step \cite{ThomasPRL}, thermal annealing \cite{HahmDefects,HarrisonInScience}, electric fields \cite{Morkved}, growth from an aligned boundary \cite{SegalmanKramer,Deepak} and imposed shear \cite{CoulonShear}.  The alignment under spreading observed by Hahm and Sibener represents a potentially new mechanism, different from these others.  To understand this mechanism is important for the overall goal of controlling copolymer film patterns.

This paper is organized as follows. In section \ref{secHomopolymer} we review the known facts about homopolymer spreading. In sections \ref{secFlowAlongTerrace} - \ref{secCopolymerSpreading} we develop our model for the copolymer film spreading. Then, in section \ref{secPSPMMASpreading} our model is compared with the experimental results on PS-PMMA, and a new dynamic mechanism of domain alignment is discussed.

\section{Spreading of Homopolymer Films }
\label{secHomopolymer}

We will start with a brief review of the spreading dynamics of regular homopolymers. This spreading has been widely investigated and is mostly understood \cite{deGennes1985, Bruinsma, Leger, deGennes1984}. The rate of this spreading, as well as the shape of the advancing liquid front, can be derived from the lubrication approximation. Within this approximation, the two-dimensional current density of material flow  $J$ is given by

\begin{equation}
J = - \frac{h^2}{\eta} \left(b + \frac{1}{3}h\right) \nabla p,
\end{equation}

\noindent
where $\eta$ is the viscosity, $h$ is drop thickness, $b$ is the slip length and $p$ is pressure inside the drop. There are generally two major contributions to this pressure: one is from Young-Laplace capillary effect $\gamma \nabla^2 h$ and the other is from Van-der-Waals disjoining pressure $W(h) = - A / ( 6 \pi h^3 )$. The coefficient $A$ is a material dependent constant. Thus

\begin{equation}
p = - \gamma \nabla^2 h + W(h).
\end{equation}

In the frame moving with velocity $U$ of the contact line, the current becomes $J' = J - U h$. For steady-state spreading, this current must vanish, and the spreading profile $h(x)$ is determined by

\begin{equation}
\label{eq_generalspreading}
\nabla \left[ - \gamma \nabla^2 h + W(h) \right] \frac{h}{\eta} \left(b + \frac{1}{3}h\right) = - U.
\end{equation}

Based on the thickness $h$ of the drop, there are three distinct regions that can be distinguished. First of all, there is a region of \textit{macroscopic drop} where film thickness is much larger than the slip length ($h \gg b$), and contribution from disjoining pressure is negligible. Equation (\ref{eq_generalspreading}) in this region reduces to the well known spreading equation investigated by Tanner \cite{deGennes1985}

\begin{equation}
\label{eq_macrodrop}
\frac{\gamma}{3 \eta} h^2 \nabla^3 h = U
\end{equation}

\noindent
which at large $h$ predicts zero curvature wedge-like profile $h \propto |x| \ln^{1/3} |x|$ and cubic dependence of the spreading rate on the apparent contact angle $U \propto \frac{\gamma}{\eta} \theta^3$. 

The second distinct region of (\ref{eq_generalspreading}) is called the \textit{precursor film} \cite{deGennes1985}, \ie the area of the film where thickness is very small so that plug flow takes place and Van-der-Waals interactions dominate. Then (\ref{eq_generalspreading}) reduces to 

\begin{equation}
\label{eq_precursor}
\frac{b h}{\eta} \nabla W(h) = \frac{A b}{2 \pi \eta} h^{-3} \nabla h = - U,
\end{equation}

\noindent
which predicts $h \propto |x|^{-1/2}$ thickness profile. 

Finally, there's an intermediate region characteristic only to polymeric drops and denoted as the \textit{foot region}. Existence of this region is associated with very large slip length of polymeric liquids, so that there is a region of the film where Van-der-Waals interactions are already insignificant, but flow is still of the ``plug'' type. Then (\ref{eq_generalspreading}) reduces to

\begin{equation}
\label{eq_foot}
\frac{b \gamma}{\eta} h \nabla^3 h = U,
\end{equation}

\noindent
with film thickness obeying \cite{deGennes1984} $h \propto |x|^{3/2}$. 

The described model for the polymer spreading sufficiently well describes both precursor and macroscopic film regions. However, $x^{3/2}$ dependence for the foot is not generally observed \cite{Leger}. 

Similar effects may be expected in copolymer drops. In the precursor film, for example, Van-der-Waals interactions with the substrates significantly exceed repulsive interactions between copolymer blocks, so that copolymer should obey the same dynamics (\ref{eq_precursor}) as a homopolymer. Similarly, in the macroscopic drop region, domain orientation may be expected to be random and macroscopic viscosity remains isotropic. Hence, the domain structure may be accounted for via a macroscopic viscosity so that general spreading dynamics still follows the Tanner's law (\ref{eq_macrodrop}). 

However, in the foot region, one may expect even stronger deviations from (\ref{eq_foot}) than those in homopolymeric liquids, as periodic domain structure of copolymer gets aligned parallel to the substrate and a quantized thickness profile emerges. Then the continuous description of equation (\ref{eq_generalspreading}) becomes invalid. To describe the flow of material inside this terraced structure we will look at two different regions of material flow: on one hand, there's continuous flow inside each terrace, which can be described from a hydrodynamical standpoint; and, on the other hand, there's activated flow at the terrace steps, where chains have to hop between discontinuous domains of different terraces.

\section{Flow Inside a Single Terrace}
\label{secFlowAlongTerrace}

\textit{Hydrodynamic Approximation.} 
The terraced structure of copolymer films comes as a result of domain alignment parallel or perpendicular to the substrate. Some of these terraces may slide on top of each other (Fig. \ref{figLayers}), with flow of material inside each terrace being strictly one or two dimensional and parallel to the substrate. Then it is convenient to characterize profile of each terrace in terms of area density $\rho \propto h$, so that the flow can be described by hydrodynamic equations of motion:

\begin{equation}
\label{eq_hydrocontinuity}
\frac{\partial \rho}{\partial t} + \nabla \cdot (\rho \vec{u}) = 0,
\end{equation}

\noindent
and

\begin{equation}
\label{eq_DarcyFull}
\rho \frac{D u_j}{D t} = \nabla_i \: \sigma_{ij} - k u_j
\end{equation}

\noindent 
where $k$ is the viscous friction opposing the flow, and $\sigma_{ij}$ is the stress tensor. In the discussion that follows, we will consider only slow ($Du/Dt \approx 0$) spreading in the $x$-direction. Additionally, we will neglect $y$-dependence of the stress tensor $\sigma$. This dependence, which arises from the domain-induced anisotropy of phase-separated copolymer and potentially affects such parameters as direction of spreading, is insignificant for the present discussion and will be briefly considered in section \ref{secPSPMMASpreading}. Hence, (\ref{eq_DarcyFull}) reduces to the familiar Darcy's law

\begin{equation}
\label{eq_navierstokes}
u_x = \frac{1}{k} \left[ \frac{\partial \sigma_{xx}}{\partial x} + \frac{\partial \sigma_{xy}}{\partial y} \right]
\approx \frac{1}{k} \frac{\partial \sigma_{xx}}{\partial x}
\end{equation}

Non-uniform stress $\sigma_{xx}$ exists in the film because it is not at its equilibrium thickness $h_0$. The stress can be expressed in terms of the free energy per chain $F_c$.  This free energy evidently has a minimum when $h = h_0$, and for $h$ near $h_0$ $F_c$ is quadratic in $(h - h_0)$ \cite{Semenov,WittenCylinders}:

\begin{equation}
\label{eq_chainfreeenergy}
F_c \approx F_0 + g_0 \gamma_{AB} v_0 \frac{(h - h_0)^2}{h_0^3} 
\end{equation}

Here $v_0$ is the volume displaced by each chain.  We treat the polymers as an incompressible liquid; thus $v_0$ is a fixed constant. The $\gamma_{AB}$ is the interfacial energy of the interface between the A and B blocks.  The quantity $g_0$ is a numerical constant that depends on the domain morphology and how it is distorted in the process of changing $h$ (e.g. $g_0 = 3/2$ for plane lamellar terraces \cite{g}).

To determine the stresses, we consider a small section of the terrace of width $\Delta x$ and length $\Delta y$ with thickness $h \neq h_0$. The section is prevented from relaxing to height $h_0$ by forces from the adjacent film.  If the section were detached from adjacent terrace, it would contract or expand laterally, allowing the height to return to its equilibrium value.  To prevent this, the adjacent terrace must exert a tensile stress on the section. In particular, during spreading in $x$ direction, $\Delta y$ stays constant and stress $\sigma_{xx}$ is

$$
\sigma_{xx} = \frac{1}{h \Delta y} \left. \frac{\partial F_T }{\partial (\Delta x)} \right|_{\Delta y} 
$$

Using incompressibility of polymer $h \Delta x \Delta y = V$ and the fact that the terrace's free energy $F_T = (V/v_0) F_c$ we finally get

\begin{equation}
\label{eq_pressure}
\sigma_{xx} 
  = - \frac{h}{v_0} \left. \frac{\partial F_c}{\partial h} \right|_{\Delta y}
  = - 2 g \frac{(h - h_0)}{h_0^2} \gamma_{AB}
\end{equation}

The numerical constant $g$ depends on the $g_0$ of Eq. \ref{eq_chainfreeenergy}, and in the trivial case of lamellar domain $g = g_0$. In general, the tensile stress also includes a Laplace-Young contribution $\gamma \nabla^2 h$ from the curvature of the free surface.  This part of the pressure is negligible in the long, flat terraces we are considering.  Then the profile of the spreading terrace is given by

\begin{equation}
\label{eq_hydrofull}
\frac{\partial h}{\partial t} =  \frac{\partial}{\partial x} \left[ \frac{2 g \gamma_{AB}}{k} \frac{h}{h_0^2} \frac{\partial h}{\partial x}  \right].
\end{equation}

Away from the terrace steps, the thickness of the terrace is close to $h_0$, so that to the first order in $h - h_0$ the thickness profile is given by

\begin{equation}
\label{eq_hydroshort}
\frac{\partial h}{\partial t} =  D_h \frac{\partial^2 h}{\partial x^2},
\end{equation}

\noindent where $D_h = 2 g \gamma_{AB} / h_0 k_{h=h_0}$.

\textit{Dissipative Forces at the Substrate.} 
We now consider the physical origin of the dissipative coefficient $k$. The most common type of viscous forces are the frictional forces at the substrate interface. These forces are most easily expressed in terms of the slip length $b$, \ie the distance below the substrate at which velocity of liquid $u(z)$ extrapolates to zero: $b = \left[ u / (du / dz) \right]_{z=0} = \eta / \beta$. Here $\eta$ is liquid viscosity, and $\beta$ is friction coefficient at the substrate. The magnitude of the friction force does not depend on terrace thickness, so that coefficient $k$ entering (\ref{eq_navierstokes}) is $k = k_s = \beta / h = \eta / b h$.

While this expression for $k_s$ is believed to be true for macroscopic drops, it fails in thin films due to rapid decrease in monomer mobility near a substrate. For instance, near a high-energy substrate monomer mobility may drop $10^4$ times compared to its bulk value \cite{Bruinsma}, leading to a similar drop in $k$. Introducing phenomenological parameter $\xi = \mu_{bulk} / \mu_{substrate}$, which relates bulk ($\mu_{bulk}$) and substrate ($\mu_{substrate}$) mobilities, the friction coefficient for the sliding chains becomes $k = \xi k_s$.

This decrease in mobility falls off rapidly away from the substrate. Then the chains that are affected most are those in immediate contact with the substrate, while remaining chains can freely diffuse between them. This leads to a two-liquid model of flow, first introduced for polymer spreading by Bruinsma \cite{Bruinsma}. According to this model, the first liquid consists of chains that come in direct contact with the substrate and slowly slide with velocity $\vec{u}_s$, while the other liquid consists of chains that diffuse between them with velocity $\vec{u}_d$. The net velocity of material flow is then

\begin{equation}
\label{eq_U}
\vec{u} = \tau \vec{u}_s + (1 - \tau) \vec{u}_d,
\end{equation}

\noindent 
where $\tau$ is the fraction of the sliding chains, \ie chains that are in contact with the substrate. If the fraction $\tau$ of sliding chains is small, we can temporarily switch to the frame moving with velocity $U_d$ where sliding chains become diffusing, with the drag force on them given by Einstein relation $\vec{f_0} = \frac{k_B T}{D_c} (\vec{u}_s - \vec{u}_d)$. Here $D_c$ is the self-diffusion coefficient for chains moving along domain walls. The net force between sliding and diffusing chains is then $\vec{f} = \tau \vec{f_0} / v_0 = \tau \frac{k_B T}{v_0 D_c} (\vec{u}_s - \vec{u}_d)$, and the Darcy's law for each of these components is

\begin{eqnarray}
(1 - \tau) \nabla \cdot \sigma + \tau k_d (\vec{u}_d - \vec{u}_s) = 0 \\
\tau \nabla \cdot \sigma + \xi k_s \vec{u}_s - \tau k_d (\vec{u}_d - \vec{u}_s) = 0
\end{eqnarray}

\noindent
where $k_d = k_B T / v_0 D_c$, $k_s = \eta / b h$ and $v_0$ is the volume of one chain. These equations, together with (\ref{eq_U}) give the final expression for the velocity:

\begin{equation}
\label{eq_slidingvelocity}
\vec{u} = - \left[ \frac{(1 - \tau)^2} {\tau} \frac{1}{k_d} + \frac{1}{\xi k_s} \right] \nabla \cdot \sigma.
\end{equation}

As expected, this expression reduces to (\ref{eq_navierstokes}) in the case when sliding dominates diffusive flow ($\xi k_s \ll k_d$). With velocity (\ref{eq_slidingvelocity}), one-dimensional film spreading is still described by (\ref{eq_hydroshort}), with collective diffusion coefficient $D_h$ being

\begin{equation}
\label{eq_Dsliding}
D_{slid} = \frac{3 \gamma_{AB}}{h_0} \left[ \frac{(1 - \tau)^2} {\tau} \frac{1}{k_d} + \frac{1}{\xi k_s} \right].
\end{equation}

\textit{Dissipative Forces Between Sliding Terraces.}
Along with dissipation at the substrate, an alternative dissipation mechanism arises when terraces slide over each other. During this sliding motion, brushes from adjacent terraces interpenetrate and viscous dissipation takes place. The areal density of the dissipated power $T\Sigma$ is given by

\begin{equation}
\label{eq_viscousdissipations}
T\Sigma = \int_{ - \zeta }^\zeta {\eta ^\ast \left( {\frac{dv}{dz}} 
\right)^2dz} \approx \eta ^\ast \frac{v^2}{ 2 \zeta },
\end{equation}

\noindent
where $v$ is spreading velocity, $\eta^{\ast }$ is the effective viscosity of one of the copolymer blocks, and $\zeta$ is the characteristic depth of interpenetration of two adjacent brushes. In the plane lamellar morphology the brush thickness $h_A \propto N_A^{2/3}$ so that the interpenetration depth is \cite{WittenInterpenetration}:

\begin{equation}
\zeta \approx \left( {R_A^4 / h_A} \right)^{1 / 3} \approx a N_A^{4/9}.
\label{eq_zeta}
\end{equation}

\noindent
where $R_A$ is radius of gyration of a free chain, and $a$ is a microscopic length scale close to the monomer size $a_0$:\cite{Semenov,WittenInterpenetration} $a \sim (kT / \gamma_{AB} a_0^2)^{1/9} a_0$. The effective viscosity $\eta^{\ast }$ entering (\ref{eq_viscousdissipations}) is qualitatively different from the homopolymer melt viscosity. This is due to the fact that chains in a segregated copolymer are bound by domain walls and may no longer undergo reptation motion characteristic of linear homopolymers. Instead, the copolymer system becomes increasingly similar to the star polymer melt, where chains are also constrained at their vertex \cite{WittenInterpenetration}. The characteristic mode of motion of star polymers is repetitive contractions and extensions of one arm of the star, with time scales exponentially increasing with the length of this arm. Correspondingly, the viscosity of star polymers also exhibits exponential increase \cite{Fetters} $\eta ^\ast \approx \eta_{arm} e^{N_{arm} / N_e}$, where $N_{e}$ is the average entanglement length, $N_{arm} = 2 \zeta N_A / h_A \sim 2 N_A^{7/9}$ is average arm length, and $\eta_{arm}$ is the viscosity of the corresponding linear polymer of length $2 N_A^{7/9}$. Thus the friction coefficient between terraces becomes

\begin{equation}
\beta = \frac{T\Sigma}{v^2} \approx \frac{\eta_{arm}}{2 a} N_A^{-4/9} \exp \left(2 N_A^{7/9} / N_e\right),
\end{equation}

\noindent
so that final dynamics of the terrace is described by (\ref{eq_hydroshort}) with collective diffusion coefficient $D_h$ given by

\begin{equation}
\label{eq_Dviscous}
D_{visc} = \frac{2 g \gamma_{AB}}{\beta} \approx \frac{6 a \gamma_{AB}}{\eta_{arm}} N_A^{4/9} \exp \left(- 2 N_A^{7/9} / N_e\right).
\end{equation}

\section{Flow At The Terrace Steps}
\label{secFlowAtTheSteps}

\begin{figure}[tb]
\centerline{\includegraphics[width=3in,height=1.2in] {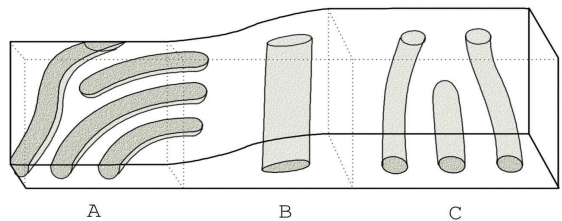}}
\caption{Schematic top view of a step between terraces with lying-down cylinders. As thiner terrace (A) grows, material from the thicker terrace (C) gets squeezed inside transitional region (B), resulting in chain hopping from squeezed bulk domains (B) towards stretched surface domains (A). For simplicity, only the topmost terraces of the film are shown.}
\label{figTerraceStepA}
\end{figure}

Besides continuous flow of material inside each of the terraces, flow of material between terraces is needed to sustain spreading. However, topologically different domains of adjacent terraces (Fig. \ref{figTerraceStepA}) are either disconnected, or connected via some intermediate morphology \cite{Harrison1998}. In either case, activated events are needed to create flow at the step: in the former case, single chain hopping is necessary, while, maintaining intermediate morphology in a dynamic system, such as spreading system of Fig. \ref{figTerraceStepA} with step moving to the right, requires many chain activation events. As the time scales for many chains events are significantly longer than for single chain hopping, from here on we will assume that single chain hopping dominates flow at terrace steps and intermediate morphologies do not form.

Qualitatively, the decrease in chain self-diffusion due to activation events has the form \cite{Barrat, Kramer}

\[
D \approx D_0 e^{ - \Delta E / kT},
\]

\noindent
where $\Delta E$ is the activation free energy for the minority block to hop through the stationary block, and $D_{0}$ is the self-diffusion coefficient of a homopolymer melt with chains assembled from majority block monomers and having the same polymerization as the copolymer chain.

There are two distinct contributions to the above activation energy: the energy cost $\Delta E_0$ for the hopping block to pass through the incompatible stationary block, slightly decreased by the excess energy, or chemical potential, of the chains inside domains. In stationary systems, chemical potentials for all domains are the same and no net flux occurs, while in the spreading systems chemical potentials may differ between terraces, producing net flow of chains.

To estimate this flow, we notice that the average rate at which chains hop between domains with chemical potential $\mu$ and spacing $h_0$ is \cite{LandauVol6}

\begin{equation}
\label{eq_deltat}
(\Delta t)^{-1} = \frac{2D}{h_0^2} \approx \frac{2 D_0}{h_0^2} e^{(- \Delta E_0 + \mu) / kT}.
\end{equation}

For two parallel lamellar layers separated by $h_0$ the current density $j_0$ of chains hopping from one layer to the other is given by $j_0  = \rho_0 (\Delta t)^{-1}$, where $\rho_0$ is equilibrium number of chains per unit area.  At equilibrium, an equal current density  flows in the opposite direction.  However, in our spreading terrace, the thin, leading terrace is not in equilibrium with the thicker terrace behind the step.  The two regions have different values of $\mu$, and a net current is expected.  

\begin{figure}[tb]
\centerline{\includegraphics[width=2.7in,height=1.1in] {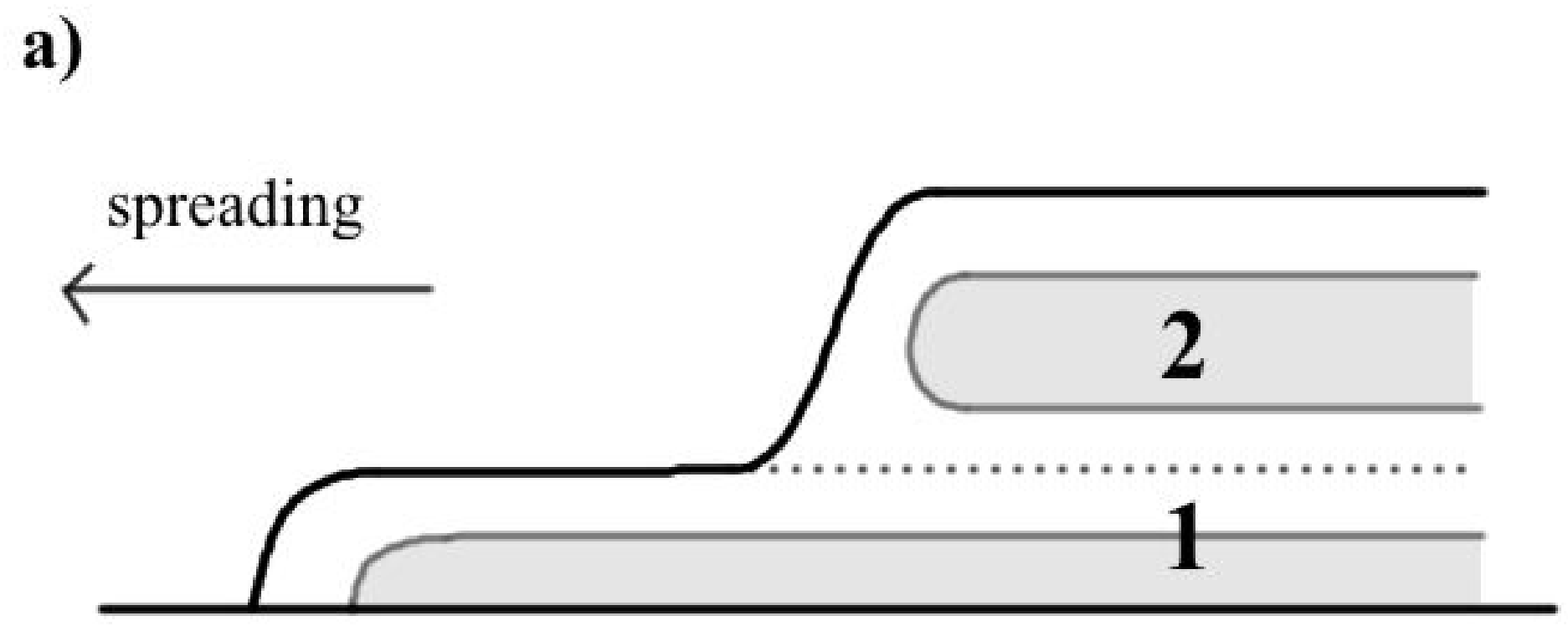}}
\centerline{\includegraphics[width=2.7in,height=1.1in] {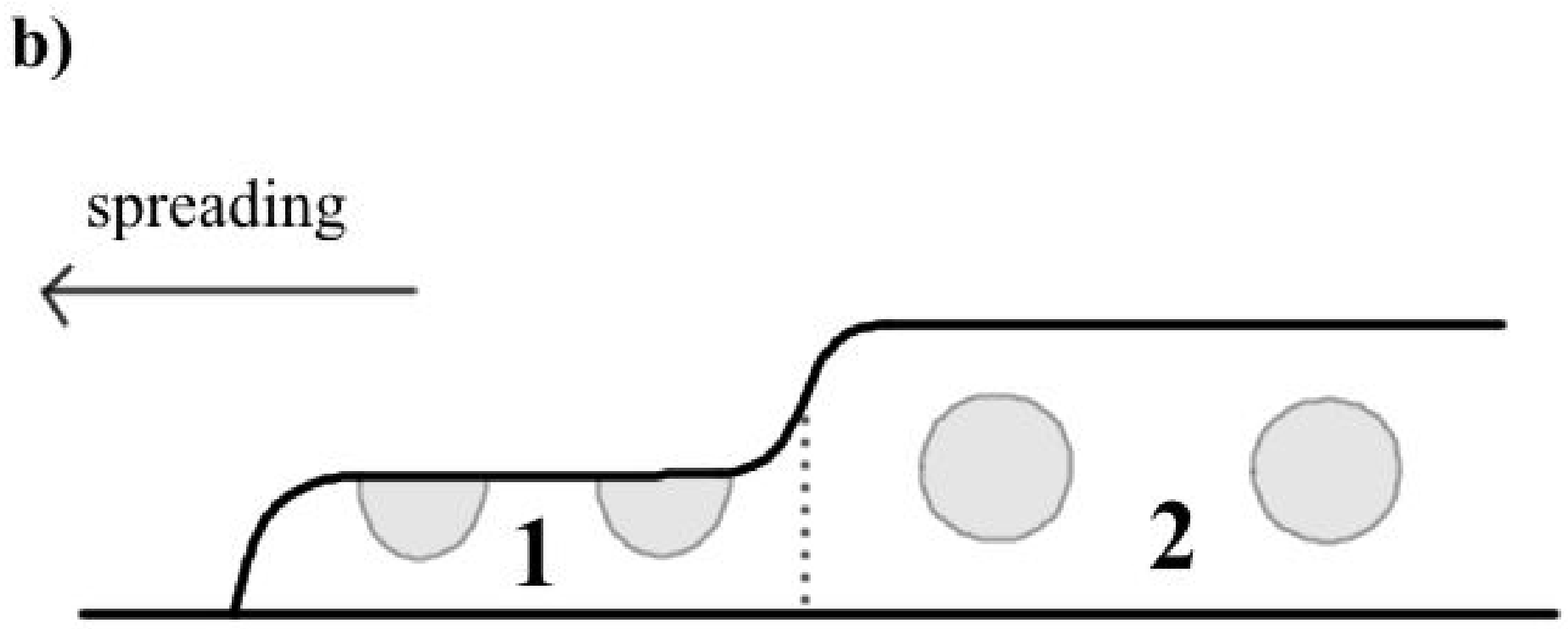}}
\caption{Two cases of stress propagation between terraces: a) if domains inside top (2) and bottom (1) terraces are separated, stress from pulled bottom terrace does not propagate into top terrace; b) if two terraces are linked (2), stress from the pulled terrace (1) equally splits between the two linked terraces (2).}
\label{figStresses}
\end{figure}

The relationship between the emitting and receiving regions may take two general forms, as depicted in Fig. 4.  In this figure the regions of different chemical potential are labeled 1 and 2. In Fig. \ref{figStresses}a the regions lie one above the other; in Fig. \ref{figStresses}b, one lies ahead of the step, the other behind it.  In view of Eq \ref{eq_deltat} the current of polymers must be largest in regions with smallest activation free energy $\Delta E$ and largest $\Delta \mu$. In the configuration \ref{figStresses}b, the change of $\mu$ is concentrated at the step and, hence, so is the current. Similarly, in configuration \ref{figStresses}a domain distortion is maximum near the step and the free energy of the transition state may be reduced. For example, the hopping distance for a B block may be reduced, and thus the amount of stretching required for a B block to make the transition should be reduced \cite{Helfand}. Thus we expect the current to be concentrated at the step in both situations.  We shall assume that this is the case in what follows.

The length of the step region where the current is concentrated has a length in the $x$ direction of order $h_0$.  The current $i$ per unit width of the step is thus approximately the net current density times $h_0$:

\begin{equation}
\label{eq_generalcurrent}
i \approx h_0 \Delta j_0 \approx \frac{2 D_0 \rho_0 e^{-\Delta E_0/kT}}{h_0} \left ( e^{\mu_h/kT} - e^{\mu_\ell/kT}\right ).
\end{equation}

The chemical potential for a terrace of width $\Delta x$ and length $\Delta y$ to the first order in $h - h_0$ is given by 

\begin{equation}
\label{eq_mu}
\mu = \left. \frac{\partial F_T}{\partial n} \right|_{\Delta x, \Delta y} \approx \mu_0 + 2 g_0 \frac{h-h_0}{h_0} \gamma_{AB}^\ast,
\end{equation}

\noindent 
where $\gamma_{AB}^\ast = \gamma_{AB} v_0 / h_0$ is interfacial energy normalized to one chain, $F_T = h \Delta x \Delta y F_c / v_0$ is terrace free energy (see Eqs. \ref{eq_chainfreeenergy} - \ref{eq_pressure}), and $n = h \Delta x \Delta v / v_0$. If terraces can slide on top of each other (as in Fig. \ref{figStresses}a), then the spreading pressure from the lower terrace does not penetrate into the higher terraces, thus leaving thicknesses and chemical potentials of the latter at equilibrium. Then $\Delta \mu = 2 g \gamma_{AB}^\ast (h - h_0) / h_0$.

On the other hand, if two terrace are bound by domain walls (as in Fig. \ref{figTerraceStepA} and Fig. \ref{figStresses}b), then spreading pressure penetrates past the terrace step. From the balance of forces at the step, the thickness $h$ of the terrace on the left (Fig. \ref{figStresses}b), and thickness $h_2$ of the double terrace on the right are related via $(h - h_0) = 2 (h_2 - h_0)$. Then $\Delta \mu = g \gamma_{AB}^\ast (h - h_0) / h_0$.

In either case, difference in chemical potentials is proportional to the deformation $h-h_0$ of the lower terrace, and net current (\ref{eq_generalcurrent}) between terraces becomes

\begin{equation}
\label{eq_stepcurrent}
i \approx - 2 \tilde{g} D_0 \rho_0 \frac{h - h_0}{h_0^2} \frac{\gamma_{AB}^\ast}{kT} e^{(-\Delta E_0 + \mu_0) / kT},
\end{equation}

\noindent
where $\tilde{g}$ is between $g_0$ and $2g_0$ depending upon the interaction between terraces.

\section{Spreading of the Copolymer Films}
\label{secCopolymerSpreading}

It was shown above that the spreading of copolymer films consists of two parts: flow of material inside each terrace, governed by diffusion law (\ref{eq_hydroshort}), and flow of material between terraces, with the flow rate described by (\ref{eq_stepcurrent}). At early stages of the spreading, when terrace lateral extension is small, one may expect the dominant contribution to the total rate to come from the flow at terrace steps, while at late times, the dominant role should shift to the flow inside terraces.

\begin{figure}[tb]
\centerline{\includegraphics[width=2.5in,height=1.4in] {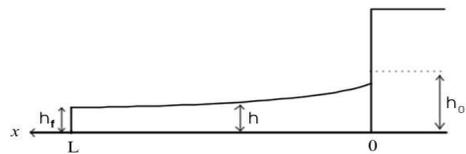}}
\caption{Profile of the advancing terrace.}
\label{figProfile}
\end{figure}

To find this time dependence, we turn to the spreading of the lowest terrace. The advancing front of this terrace ($x = L(t)$, Fig. \ref{figProfile}) is assumed to be pulled with a constant pressure from the wetting forces. As pressure and thickness of a terrace are proportional (eq. (\ref{eq_pressure})), this is equivalent to stating that advancing front is held at a constant thickness $h_f$:

\begin{equation}
\left. h \right|_{x=L(t)} = h_f.
\label{eq_boundary1}
\end{equation}

The boundary condition at the step between terraces ($x = 0$, Fig. \ref{figProfile}) arises from the balance between the current (\ref{eq_stepcurrent}) entering from the higher terrace, and the diffusion current $i = - \frac{\rho_0}{h_0} D \frac{\partial h}{\partial x}$ inside the terrace, where $D$ is collective diffusion constant entering (\ref{eq_hydroshort}) and defined by (\ref{eq_Dsliding}) or (\ref{eq_Dviscous}). Hence

\begin{equation}
\label{eq_boundary2}
- \frac{\partial h}{\partial x} + A (h - h_0) = 0, {\rm     at } x = 0,
\end{equation}

\noindent
where

\begin{equation}
\label{eq_A}
A = 2 \tilde{g} \frac{D_0}{D} \frac{\gamma_{AB}^\ast}{kT} \frac{1}{h_0} e^{(-\Delta E_0 + \mu_0) / kT}.
\end{equation}

Finally, the lateral extension $L$ of the terrace is related to the current $i_f$ at the advancing front via $i_f = \rho_0 (h_f/h_0) (dL/dt)$. Hence

\begin{equation}
\frac{dL}{dt} = - \frac{D}{h_f} \left. \frac{\partial h}{\partial x} \right|_{x=L(t)},
\label{eq_dLdt}
\end{equation}

\noindent
with

\begin{equation}
\left. L \right|_{t=0} = 0.
\label{eq_boundaryL}
\end{equation}

Thus we come to a system of equations (\ref{eq_hydroshort}, \ref{eq_boundary1}, \ref{eq_boundary2}, \ref{eq_dLdt}, \ref{eq_boundaryL}). Exact solution of this system is not trivial, so we will look only on large and small time asymptotics. 

The large time behavior is best analyzed in new coordinate $z = x / L(t)$, which eliminates the moving boundary condition (\ref{eq_boundary1}). Differential equation (\ref{eq_hydroshort}) with moving boundary condition, similarly to the case with stationary boundaries, has only one solution \cite{Tihonov}. Then, in the $z$ coordinate, the system is likely to asymptotically approach some stable state $h_1(z)$. Assuming that $h(z, t) = h_1(z) - h_2(z) t^{-\beta}$, $L = a t^\alpha$, we get $\alpha = \beta = 1/2$, and

\begin{equation}
h_1(z) = h_f + (h_0 - h_f) \left[ 1 - \frac{\erf{R z}}{\erf{R}}\right],
\end{equation}

\begin{equation}
h_2(z) = (h_0 - h_f) \frac{\exp{(-R^2 z^2)}}{A \sqrt{\pi D} \ \erf{R}} \left[1 - \frac{\erfi{R z}}{\erfi R} \right],
\end{equation}

\begin{equation}
\label{eq_spreadingdistance}
L = \sqrt{2 D \frac{h_0 - h_f}{h_f}} \ t^{1/2},
\end{equation}

\noindent
where $R = \sqrt{\frac{h_0 - h_f}{2 h_f}}$. Thus velocity of spreading at late times is given by

\begin{equation}
v_{{\rm late}} = \sqrt{\frac{D}{2} \frac{h_0 - h_f}{h_f}} \ t^{-1/2}.
\label{eq_latevelocity}
\end{equation}

The early times behavior, on the other hand, is characterized by nearly constant thickness $h \approx h_f$ of the growing terrace (Fig. \ref{figProfile}). Then all the material entering a terrace (eqs. \ref{eq_stepcurrent} and \ref{eq_boundary2}) should go towards lateral extension of that terrace with spreading velocity given by

\begin{eqnarray}
\label{eq_earlyvelocity}
v_{\rm early} & = & \frac{dL}{dt} \approx - i \frac{h_0}{\rho_0 h_f} \\ \nonumber
  & \approx & 2 \tilde{g} D_0 \frac{h_0 - h_f}{h_0 h_f} \frac{\gamma_{AB}^\ast}{kT} e^{(-\Delta E_0 + \mu_0)/ kT}.
\end{eqnarray}

The characteristic cross-over time between the two velocities is found by setting $v_{\rm early} \approx v_{\rm late}$:

\begin{equation}
t_c \approx \frac{1}{2 D A^2} \frac{h_f}{h_0 - h_f}.
\label{eq_crossover}
\end{equation}

\section{Spreading of PS-PMMA Films on SiO$_2$ and Domain Alignment}
\label{secPSPMMASpreading}

\textit{Spreading rate of PS-PMMA.}
The theory above predicts the absolute spreading rates and thickness profile to be expected in the experiments of Ref \onlinecite{Hahm}.  We may thus test our theory by comparing these expected features with the observed ones. In these experiments, a thin film of 80,000 amu PS-PMMA was spin cast onto a silicon substrate that was initially covered with a minor polar solvent. The presence of the solvent resulted in the formation of annuli structures, likely caused by the dewetting of the substrate, and consisting of a tall rim, bare substrate inside the rim, and uniform film outside (Fig. \ref{figHahm}). Later, as the solvent evaporated, the material started flowing downwards onto the bare substrate area inside the rim. This flow is presumed to be driven by wetting forces, so that its rate should be expressed by (\ref{eq_latevelocity}) - (\ref{eq_earlyvelocity}).

The structure of asymmetric, cylinder-forming PS-PMMA copolymers on SiO$_2$ used in this experiment is well known \cite{Jaeger,Hahm,HahmDefects}, and consists of plane lamellar layer on the substrate, followed by a series of layers (terraces) with cylindrical domains inside (Figure \ref{figLayers}). To estimate the spreading rate of this system, we turn to the bottommost terrace pulled by the wetting forces. The collective diffusion constant for material flow inside this terrace is given by (\ref{eq_Dsliding}) and contains two separate contributions. On one hand, there's a friction on the terrace as a whole (second term in Eq. \ref{eq_Dsliding}). However, the hydrogen bonding between PMMA and SiO$_2$ is very strong ($\xi \gg 1$), and contribution from this term to the diffusion constant is small. The first term, on the other hand, describes self-diffusion of fast chains through the matrix of substrate-bound slow chains. Since there are no walls obstructing this flow in the bottommost terrace of PS-PMMA, contribution from this term dominates and $D_{slid} \approx 3 (1-\tau)^2 v_0 D_c \gamma_{AB} / k_B T \tau h_0$.

The fraction $\tau$ of chains that are in contact with the substrate can be estimated based on a strongly-stretched approximation \cite{Semenov}, which predicts that the density of chain ends in a flat brush is $g(r) \propto r / (h^2 - r^2)^{1/2}$. Since strongly stretched approximation neglects deviations of chain paths from their mean, and all paths are presumed to go unidirectionally towards the free end of the brush, we can assume that all chains ending within a certain small distance from the substrate are bound by the substrate. We estimate this distance as the penetration length $\zeta$ discussed in (\ref{eq_zeta}). This is the length scale within which the polymers behave like unperturbed random walks rather than like strongly stretched chains. For a $N_A = 200$ PMMA chains with radius of gyration $R_A \approx a_0 (N_A/6)^{1/2}$, we find $\zeta \approx (R_A/h_A)^{4/3} h_A \approx .5\;h_A$.  Then $\tau$ can be found via

\begin{equation}
\tau \approx \left( {\int_{h_A-\zeta}^{h_A} {g(r)dr}} \right) / \left( \int_{0}^{h_A} {g(r)dr} \right) \approx {\rm 85\%}.
\end{equation}

In fact, this is an upper estimate for $\tau$, as density of ends $g(r)$ has a singularity near the open end of the brush produced by the neglect of entropy of chain paths. The lower estimate for $\tau$ would come from the assumption of constant $g(r)$, so that $\tau \geq a_0 / h_A \sim 10\%$, with actual value being somewhere in between. Fortunately, this variation in $\tau$ produces only an insignificant variation in the prefactor of $D_{slid}$. 

Finally, the self-diffusion constant for the chain flow along domain walls $D_c$ is also not readily available. Still, its value can be estimated from the analogy between copolymer flow along domain walls and flow of star polymers, whose self-diffusion is given by $D_{star} \approx D_{arm} \exp(-N_{arm}/N_e)$. Here $N_e \approx 180$ is average number of monomers per entanglement, and $D_{arm}$ is self-diffusion of linear polymers forming each arm. Then the self-diffusion coefficient for a PS star polymer with 60,000 amu arms is  $D_{\rm PS} \approx .4 \cdot 10^{-15}$  m$^2$/s, and the self-diffusion coefficient for a PMMA star polymer with 20,000 amu arms is $D_{\rm PMMA} \approx .4 \cdot 10^{-15}$  m$^2$/s. \cite{Watanabe,Milhaupt} Because of similarity of these two coefficients, it is reasonable to assume that the self-diffusion coefficient for the whole block copolymer moving along domain walls is also $D_c \approx .4 \cdot 10^{-15}$  m$^2$/s. Then the forced collective diffusion coefficient for the flow along bottommost terrace (\ref{eq_Dsliding}) is $D_{slid} \approx .7 \cdot 10^{-15}$ m$^2$/s, where known values for other parameters have been used \cite{Wu,Hahm,HelfandTagami}: $\gamma_{AB} \approx .6 \cdot 10^{-3} {\rm N/m}$, $h_0 \approx 22$ nm, and $v_0 \approx N a^3 \approx 1.7 \cdot 10^{-25}$ m$^3$.

The final parameter influencing spreading dynamics (\ref{eq_A}) - (\ref{eq_earlyvelocity}) is the activation free energy $\Delta E_0$. Its magnitude strongly depends on the polymer conformation in the transition state. Two distinctly different conformations are possible \cite{Helfand}. In the first conformation, the hopping block propagates as a compact blob, thus minimizing its stretching energy at the expense of larger interfacial energy. In the second conformation, the hopping block is stretched between original and destination domains, so that only a fraction of the hopping monomers comes in contact with the incompatible stationary block. Using the calculations by Helfand \cite{Helfand}, one can show that stretching mechanism dominates in the PS-PMMA films under discussion, and activation energy is roughly $\Delta E \approx \chi N_{PMMA} kT$. 

Based on these quantities, we can now estimate the spreading rates of the PS-PMMA film.  We made these estimates without any assumptions about the spreading pressure.  This pressure depends on the molecular structure and interfacial shape on an atomic scale at the advancing front.  It is difficult to estimate it without an accurate characterization of the surface and the monomers at this scale.  Accordingly, we used the observed thickness at the front, namely $h_f$ instead of the pressure.  It is observed to be about $0.85$ times the equilibrium terrace thickness $h_0$.  Using this knowledge, we can apply Eqs \ref{eq_spreadingdistance} - \ref{eq_crossover} directly.

First of all, the crossover time $t_c$ from the step-limited to the diffusion-limited dynamics, using Eq. \ref{eq_crossover}, is found to be about .25 hours.  Hence diffusive flow along the terrace should dominate the spreading dynamics during most of the 24 hours of the experiment reported in Ref \onlinecite{Hahm}.  Second, the typical distance $L$ over which PS-PMMA film spreads in 24 hours at 523 K is predicted to be about 5 microns (Eq. \ref{eq_spreadingdistance}).  This is indeed close to the observations by Hahm and Sibener \cite{Hahm}, where rim width was observed to increase by 1 -- 3 microns.  Finally, the spreading velocity after 24 hours is about 100 nm/hr.

\textit{Domain alignment.}
It was already mentioned that hopping of chains between domains at terrace steps may account for alignment of the cylindrical domains in a spreading film. Specifically, in thin films, domain orientation is influenced by the nearby substrate, with domains usually oriented parallel or perpendicular to the substrate. For example, in PS-PMMA films on SiO$_2$ lying-down cylinders are formed on $nH-$thick terraces (where $H$ is domain period), while vertical cylinders are formed on $(n+1/2)H-$thick terraces. Spreading of the film and lateral growth of the terraces should then be accompanied by changes in the observed domain patterns. Specifically, here we will discuss the growth and alignment of the lying-down cylindrical domains on the growing terraces.

We first look at domain growth at the steps with horizontal domains on both sides of the step (Fig. \ref{figTerraceStepA}). This is the case, for example, at the ''double steps`` in the aforementioned PS-PMMA systems. Here term ``double step'' refers to the steps between $nH-$ and $(n+1)H-$thick terraces. 

Following section \ref{secFlowAtTheSteps}, topologically different domains of adjacent terraces may be assumed disconnected from each other (Fig. \ref{figTerraceStepA}) and the flow of material between terraces is governed by single chain hopping events. The direction of this flow is determined by the difference in terrace chemical potentials, and during spreading goes towards stretched lower terrace. Then, upon arrival to the growing terrace, a copolymer chain has a strong energetic bias to merge into existing domains rather than nucleate new ones. Hence the spreading should lead to the extension of the existing domains ensuring growth of continuous, defect free domain patterns.

\begin{figure}[tb]
\centerline{\includegraphics[width=2.5in,height=1.4in] {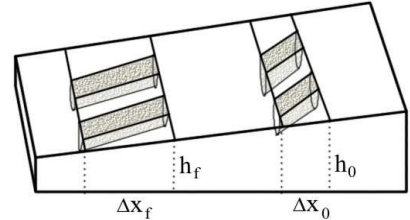}}
\caption{Change of domain orientation during film thinning. The spreading of the film to the left carries a given region A in the step to the thinner region B, thereby reducing the angle $\theta_0$. }
\label{figChangeOfAngle}
\end{figure}

The thickness gradient in the step region, and to a smaller extent in the bulk of the terrace, can also account for the \textit{alignment} of the growing surface domains in the thickness gradient direction. To demonstrate this alignment effect, we first imagine that the cylindrical surface domains meet the transitional region at an oblique angle $\theta_{0}$ to the growth direction, as shown in Figure \ref{figTerraceStepA}. Since surface domains are not linked to the bulk domains of the thicker terrace, they have to end within the transitional region where the thickness is somewhat greater than that of the lower terrace. Then a small increment of motion of this system consists of two separate processes: first, a small amount of new material is added to the ends of the surface domains; second, these domains, being part of the growing lower terrace, move in the direction of the thickness gradient. Film thinning combined with conservation of volume then leads to a change in the direction of these domains, as described below.

The growth increment extends the domains. This growth, we presume, occurs without domain bending so that $\theta_{0}$ remains unchanged, leaving the domain in the lowest energy unbent state. The translation of the step causes a given element of the film to compress in thickness and to lengthen in the direction of motion in order to conserve its volume (Fig. \ref{figChangeOfAngle}). The resulting deformation resembles that of a piece of dough being rolled under a rolling pin. As domains thin, their angle $\theta_{0}$ decreases. To quantify this decrease we notice that change in thickness of the film from $h_{0}$ to $h_{f}$ leads to its lateral extension $\Delta x_f / \Delta x_0 = h_0 / h_f$ along the thickness gradient direction $x$ (Fig. \ref{figChangeOfAngle}), while lateral extension $\Delta y$ along the step edge remains unchanged. Then the flattened domains emerge at an angle $\theta_{f}$ with the step edge, which is given by $\tan \theta _f = \Delta y_f / \Delta x_f = \tan \theta _0 \cdot h_0 / h_f $, and the extruded domains are bent towards the normal of the interface.

This mechanism alone causes domains growing in an arbitrary direction to emerge partially aligned to the normal. Further effects should cause this alignment to increase progressively with time. To demonstrate such an effect, we imagine that domains grow in increments of length $s$. During each of these increments domains extend by distance $s$, and then propagate the same distance along the gradient direction. The change in thickness that accompanies this motion, results in bending of domain tip from angle $\theta_{0}$ to $\theta_{s}$ with latter given by $\tan \theta _s = \tan \theta _0 \cdot h_s / h_0 = \tan \theta _0 \left( {1 - \frac{\left| {\Delta h} \right|}{h}s} \right)$. The relaxation of these bent domains into straight ones would require collective motion of large number of chains and exponentially long times. As this is highly unlikely, the next step of the growth will start with domain ends oriented at a smaller angle $\theta_{s}$ to the normal, leading to progressive decrease in $\theta_s$ with each growth step:

\[
\frac{\Delta \tan \theta }{\Delta s} = - \frac{\left| {\nabla h} 
\right|}{h}\tan \theta ,
\]

\noindent
and, hence,

\begin{equation}
\tan \theta = \tan \theta _0 \;e^{ - \frac{\left| {\nabla h} \right|}{h}s}.
\label{eq_anglerelaxation}
\end{equation}

Thus the growing domains progressively align in the normal to the terrace step direction. The characteristic length for this effect is determined by the size of the step region $h / |\nabla h| \sim 2 - 4 \: h$ so that characteristic length for alignment is several domain widths.

\begin{figure}[tb]
\centerline{\includegraphics[width=3in,height=1.2in] {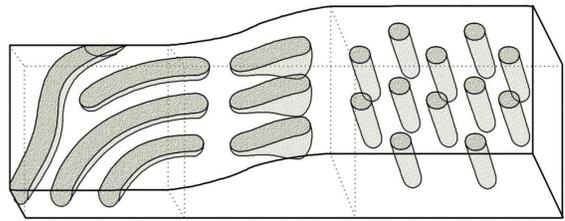}}
\caption{Schematic top view of a step between terraces with lying-down semicylindrical domains on the growing lower terrace and standing cylinders on the shrinking thicker terrace. Only the topmost terraces of the film are shown.}
\label{figTerraceStepB}
\end{figure}

To complete this section, we turn to an alternative combination of domain morphologies at terrace steps, with lying-down semicylindrical domains on the growing lower terrace and standing cylinders on the shrinking thicker terrace (Fig. \ref{figTerraceStepB}). In particular, this setup is often observed in PS-PMMA films \cite{Hahm,Jaeger}. While the general dynamics of the spreading should remain the same, some aspects of it may differ. First of all, the flow of material between terraces may proceed faster as some standing cylinders of the thicker terrace may continuously deform into short lying-down cylindrical domains (Fig. \ref{figTerraceStepB}) that later merge with the existing domains of the lower terrace. While this effect should have little effect on the net rate of spreading, which is primarily determined by collective diffusion in the bottommost terrace (not shown in Fig. \ref{figTerraceStepB}), it may lead to faster disappearance of the terraces with standing cylindrical domains, in agreement with experimental observations \cite{Hahm}.

Another difference comes from the hexagonal packing of the vertical domains, as it creates preferred directions for easy merging of vertical domains into horizontal domains. Since these directions do not depend on the wetting forces, there may emerge a misalignment between direction of spreading and direction of terrace growth, which was indeed observed in experiments on annuli structures \cite{Hahm}.

\section{Discussion}
\label{secDiscussion}

We have shown that layered structure of spreading copolymer films should lead to different type of dynamics compared to macroscopic homopolymer or simple liquid drops. First of all, the spreading of copolymer films is determined by combination of two factors: the rate of chains flow inside constant-thickness terraces of the film and rate of chain flow between terraces. The hydrodynamic flow of chains along the terraces leads to the diffusion type law for spreading (\ref{eq_hydroshort}), and its contribution dominates at late times of spreading, with velocity decreasing as $v \propto t^{-1/2}$. The hopping mechanism of chain flow at terrace steps, on the other hand, controls the early stages of spreading. The crossover time between the two regimes is given by (\ref{eq_crossover}). It is notable that both of these regimes are different from the spreading observed in macroscopic drops, where spreading rate decreases as $v \propto t^{-0.9}$. However, the $v \propto t^{-1/2}$ behavior is similar to the generic spreading rate of terraced fluids predicted be de Gennes and Cazabat \cite{Cazabat}.

Both regimes of spreading are predicted to exhibit strong dependence on chain polymerization. At early times (eq. \ref{eq_earlyvelocity}), the dominant contribution to the spreading rate comes from the activation free energy $\Delta E_0 \propto N^\alpha$, so that $v \propto N^{-2.7} e^{-N^\alpha / N_e}$, as opposed to $v \propto \eta^{-1} \propto N^{-3}$ in macroscopic homopolymer drops. Here $\alpha = 1$ for chains that hop between domains in a single blob, and $\alpha = 2/3$ for chains that stretch between domains during hopping \cite{Helfand}.

Dependence on $N$ becomes even more complicated at large times, when diffusion along terraces dominates. It was mentioned above that the spreading rate at large times is primarily determined by chain flow in the bottommost terrace with collective diffusion constant $D_{slid}$ (\ref{eq_Dsliding}). For most copolymers, the main contribution is expected from the first term in (\ref{eq_Dsliding}), which describes collective diffusion of fast chains through the matrix of substrate-bound slow chains. Then $D_{slid} \propto (h_0 \tau k_d)^{-1} \propto N^{-4/3} e^{-N / 2N_e}$. However, in systems with large degree of polymerization or small decrease in mobility $\xi$ near substrate, the second term becomes dominant with spreading rate independent of $N$: $D_{slid} \propto (h_0 k_s)^{-1} \propto N^0$. Similarly, spreading rate should be independent of $N$ if domains in the bottommost terrace obstruct continuous flow of chains, so that two-liquid approximation is inapplicable and first term in (\ref{eq_Dsliding}) disappears.

The flow of material between domains creates a dynamic process of domain growth at the terrace steps. The chains entering the growing terrace preferentially merge the existing domains in that terrace, leading to their continuous growth. In the previous section we have argued that this growth has preferential directions leading to the dynamic alignment of the domain. Before proceeding with further discussion of this model, it seems useful to compare it with other potential mechanisms for domain alignment, such as static, stress-induced alignments. 

In the annuli structures described above, stresses may come from the viscous flow, from the terrace steps, or may be caused by the change in circumference as material flows towards the center. In order to alter domain morphology, these stresses must be comparable to the internal stresses in the film created by domain walls. In films with thickness $h$ and surface tension between block $\gamma_{AB}$, the internal stresses are roughly $\sigma_{wall} \sim \gamma_{AB} / h \sim .6 \cdot 10^5$ N/m$^2$. 

This is significantly larger than the viscous stress produced by the interpenetrating brushes $\sigma_{viscous} \sim \eta^\ast v / 2\varsigma \sim 2 \cdot 10^{-2}$ N/m$^2$ (compare with (\ref{eq_viscousdissipations})), making viscous stress an unlikely candidate for the cause of observed alignment.

The other two stresses, on the other hand, come within one order of internal stresses, and could play some role in the alignment. Specifically, stress from a contact line is $\sigma_{wetting} \sim \gamma (h_0 - h_f) / h_0^2 \sim 10^4$ N/m$^2$ (where $h_h$ and $h_f$ are thicknesses of relaxed and pulled films), and stress from the change in circumference is $\sigma_{c} \sim (\gamma / h_0) (\Delta R / R) \sim 10^4$ N/m$^2$ (where $R$ and $\Delta R$ are radius and width of the rim). Despite the similarity of these stresses to the internal stresses in the film, the alignment caused by them would progressively improve towards the center of the rim, which contradicts the observations on annuli structures \cite{Hahm}.

The dynamic model of domain growth presented above, on the other hand, predicts better alignment on the grown parts of the terraces, that is in the central parts of the terraces and near the thicker terraces, in agreement with the experiment. This suggests that produced alignment is indeed caused by the dynamic effects accompanying spreading. 

Our dynamical model of domain growth may provide a new way of producing aligned domain patterns. In particular, any motion of terrace steps in the phase-separated film should lead to the growth of aligned domains on the growing terraces. One such experiment, besides the aforementioned experiment with annuli structures, is reported in Ref \onlinecite{Deepak}. In this experiment, copolymer is deposited onto a corrugated grating rather than a flat substrate. After the solvent has evaporated, the copolymer starts climbing up along the groove walls of the grating leaving aligned domain patterns inside the grooves. As this growth proceeds on time scales much longer than the thickness quantization, the observed alignment may be a result of the terrace growth.

An alternative way to stimulate terrace growth may come via control over initial nucleation of islands and holes in the film. Specifically, under normal conditions, this nucleation happens at random places. However, if one applies periodic external catalyst of the island formation, such as periodic temperature gradient, periodic electric field, or, possibly, a standing sound wave, then the formation and growth of islands and holes may proceed in controlled manner leading to the growth of aligned domains.

\section{Conclusion}

We have shown that spreading of the phase separated copolymer films is different from the spreading of homopolymer and other simple liquids. This difference is driven by domain walls inside copolymer and formation of quantized thickness terraces. As a result, regular isotropic reptation mechanism is replaced by anisotropic flow of chains along domains, and slow activated flow between domains. We have shown that at early stages, activated flow between domains at the terrace steps plays a major role, giving constant velocity of spreading. At the late times, hydrodynamic flow inside terraces dominates, resulting in diffusion type velocity decrease $v \propto t^{-1/2}$, similar to the generic spreading of terraced fluids predicted be DeGennes and Cazabat \cite{Cazabat}.

The hopping nature and slow rate of chain flow at the terrace steps were shown to produce aligned and continuous domain patterns on the  growing lower terraces. Specifically, as new chains enter the growing terrace, they tend to merge with existing domains thus leading to their continuous extension. Furthermore, thickness gradient in the area of domain growth, \ie near terrace steps, was shown to align growing domains perpendicular to the steps between terraces, thus leading to formation of aligned domain patterns on the growing terrace.

This work shows that wetting flows offer a powerful means of controlling domains in copolymer films. The effect seems to arise not from flow-induced stress, but rather by flow-induced forced hopping of chains at terrace boundaries. The alignment arises at these boundaries, as in the conventional zone refining of crystals. As suggested above, one may imagine many ways to induce wetting flows in order to achieve a wide variety of morphologies. The potential for this new structuring mechanism remains to be explored.

\section*{Acknowledgments}

We gratefully acknowledge Prof. Steven Sibener, Deepak Sundrani and Wendy Zhang for their valuable discussions. This work was supported in part by the National Science Foundation under Award Number DMR-9975533 and in part by its MRSEC Program under Award Number DMR-980859.


\begin{thebibliography}{99}
\bibitem{g} Within strongly-stretched approximation \cite{Semenov,WittenCylinders}, an exact expression for the chain free energy in lamellar domains is $F = (1/2 h^2 / h_0^2 + h_0/h) \gamma^\ast_{AB} \approx 3/2 \gamma^\ast_{AB} (h-h_0)^2 / h_0^2$, so that $g = 3/2$.
\end{thebibliography}
\end{document}